\documentclass[smallextended]{svjour3}                     
\smartqed  
\usepackage{graphicx}
\usepackage{amsmath}
\journalname{Wood Science and Technology}

\begin{document}

\title{Adhesive penetration in Beech wood}
\subtitle{Part II: Penetration Model}
\author{ Miller Mendoza \and Philipp Hass \and Falk K. Wittel \and Peter Niemz \and Hans J. Herrmann }
\institute{\at    Institute for Building Materials \\              ETH Zurich \\              Schafmattsr. 6\\              CH-8093 Zurich, Switzerland\\              }
\date{Received: date / Accepted: date}
\maketitle

\begin{abstract}
  We propose an analytical model to predict the adhesives penetration
  into hard wood. Penetration of hard wood is dominated by the vessel
  network which prohibits porous medium approximations. Our model
  considers two scales: a one dimensional capillary fluid transport of
  a hardening adhesive through a single, straight vessel with
  diffusion of solvent through its walls and a mesoscopic scale based
  on topological characteristics of the vessel network, where results
  from the single vessel scale are mapped onto a periodic network.
  Given an initial amount of adhesive and applied bonding pressure, we
  calculate the portion of the filled structure. The model is applied
  to beech wood samples joined with three different types of adhesive
  (PUR, UF, PVAC) under various growth ring angles. We evaluate
  adhesive properties and bond line morphologies described in part I
  of this work. The model contains one free parameter that can be
  adjusted in order to fit the experimental data.

\end{abstract}
\section{Introduction}\label{intro}
The most important principle in timber engineering to produce
structural wood components of constant quality, consists of cutting
wood into smaller pieces, selecting the best ones, and joining them
again by adhesive bondings. What is known as a rather simple
processing step becomes quite complicated, once we look in detail at
the penetration of the hardening adhesive into the porous wood
skeleton. Unfortunately the details of the adhesive penetration can
influence bond performance in multiple ways and the quality of the
adhesive bonds determine the overall performance of structural parts
\cite{marra-92,Custodio-etal-2009}. What complicates studies of
adhesive penetration is the interplay between pore space geometry and
fluid transport, cell wall material and adhesive rheology and of
course process parameters like amount of adhesive, growth ring
orientation, and surface roughness, just to name a few
\cite{kamke-lee-2007}. While for soft wood predictions are rather
simple, the micro-structure of hard woods complicates the problem
significantly, since adhesives can penetrate through the big vessel
network deep into the wood structure \cite{siau-84}. In a previous
work we explored the topological characteristics of the vessel network
in beech \textit{(Fagus sylvatica L.)} \cite{philipp-09} and showed in
part I of this work how the problem is dominated by flow through the
vessel network.

Adhesive penetration into hard wood was studied before
\cite{wang-yan-2005,kamke-lee-2007,sernek-resnik-kamke-99,niemz-etal-2004,collett-72},
although only experimentally or in descriptive form. For soft wood,
the penetration depth can be expressed by a simple trigonometric
function, describing the filling of cut tracheids \cite{suchsland-58}.
For hard wood however a model that characterizes the wood anatomy in
order to predict the penetration depth and the amount of adhesive
inside the structure is unknown. We construct an analitycal model
based on the network properties and predict the adhesive penetration
and the saturation of the vessel pore space. Our model has two-scales:
the first scale describing the transport of a hardening adhesive
through a single vessel in time due to an applied pressure and
capillarity effects, and also with the possibility of constant
diffusion of solvent through the vessel wall, what turns out to be
important for some adhesives like PVAC or UF. When the viscosity
increases by hardening and/or loss of solvent, the adhesive front
slows down and finally stops. On the second, or network scale, the
result for single vessels is embedded into a network model with
identical topological properties like pore size distribution and
connectivity that are characteristic for the vessel network of the
respective wood. The model is compared with experiments where
specimens are bonded with parallel longitudinal axes under varying
growth ring angles using three different adhesive systems: PRF, UF,
and PVAC. First we describe the rheological model of the adhesives,
before we calculate the penetration into a single vessel with
diffusion into the half space. Subsequently we discuss the network
construction and the consideration of process parameters. With all
model components at our hands, we finally compare the model with the
experiments and discuss the results.
\section{Model Description}\label{modeldescription}
Adhesive penetration is the result of an interplay of adhesive
hardening, capillary penetration, and technological processing. In
order to set up a model for adhesive penetration of hard wood, we have
to combine several models in a hierarchical way. First we address bulk
viscosity evolution of adhesives due to generic hardening mechanisms.
On the fundamental level, we model the penetration of a fluid into one
single, straight or wavy pipe. This model is enriched by diffusive
transport of solvent through its wall. On the next hierarchic level,
we project the fundamental model onto a network structure of perfectly
aligned hard wood that represents the vessel network. Finally, we
rotate the result of the vessel network penetration to consider the
general situation, where the adhesive surface is not necessarily
aligned to the material orientation. We show how material parameters
like porosity, hardening time or applied amount of adhesive will limit
penetration.
\subsection{Modeling the hardening process}
The hardening process of various adhesives can be described by the
temporal evolution of the viscosity $\eta$. Depending on the hardening
type, different viscosity models need to be applied. For example
reactive adhesives do not depend on the solvent concentration, while
the viscosity evolution of solvent based adhesives strongly depends on
solvent concentration $C$. In part I \cite{part1} of this work, we
showed experimental viscosity measurements for UF, PVAC, and PUR. If
solvent concentrations are important, like in the case of PVAC, the
viscosity evolution can be expressed by
\begin{equation}\label{nuall}
  \eta(C, t)=\eta_g (C) [1+\gamma(C) \exp(\alpha(C) t)] \exp(\beta [1-C]) \quad ,
\end{equation}
where $\eta_g$, $\gamma$, $\alpha$ and $\beta$ are parameters that
depend on the adhesive type and the initial solvent concentration. For
PVAC adhesive, we find $\gamma$$=$$\alpha$$=$$0$, since the hardening
process is mostly due to the loss of moisture and the initial
viscosity only depends on the initial concentration. For PUR adhesive,
the same expression can be used, however the concentration is kept
constant during the process, expressed by $\beta$$=$$0$ and constant
$\eta_g$, $\gamma$, $\alpha$ that only depend on the initial
concentration.

Unfortunately a whole class of adhesives, cannot be described by
Eq.~\ref{nuall}, since their hardening process is more complex. For
example the UF adhesive changes from liquid phase to gel phase during
penetration, resulting in penetration arrest. The only active
processes after this phase transition are the chemical curing
reactions. Therefore the viscosity model should take into account the
critical time when the phase transition occurs. Additionally, the
concentration of the solvent changes in time due to the diffusion of
the solvent into the cell wood structure. We propose the viscosity
relation
\begin{equation}\label{nuallUF}
  \eta_{\rm UF}(C, t)= d_1 \frac{\exp\left( b_2 \left(1-\exp\left(-\frac{a_1}{t}\right) \right)\right)}{c_1 - t} \quad ,
\end{equation}
where $d_1$, $b_2$, $a_1$ and $c_1$ are experimental parameters. Using
the data from Ref.~\cite{part1} we found $d_1$$=$$6.985\times 10^4$
mPa$\cdot$s$^2$ and $a_1$$=$$10510$s and variable parameters ($b_2$
and $c_1$) that depend on the initial solvent concentration. Note that
$c_1$ describes the time when the penetration process finishes due to
the liquid-gel transition. Using these two generic hardening models,
we are able to describe the viscosity evolution of numerous adhesives.
\subsection{Single vessel penetration}\label{singlechannel}
The fundamental scale is given by the capillary transport of a fluid
characterized by its viscosity $\eta$, inside a cylindrical pipe of
radius $R$ \cite{washburn-21} with a penetration rate $dl/dt$ that
follows
\begin{equation}\label{velocitypene}
  \frac{dl}{dt}=\frac{\mu}{8\eta l}R^2 \quad ,
\end{equation}
where $\mu=P_A+2\sigma \cos(\theta)/R$ with the applied pressure
$P_A$, the surface tension $\sigma$ and the contact angle between
fluid and pipe wall $\theta$. To obtain the penetrated distance $l(t)$
we integrate
\begin{equation}\label{velocitypeneint} 
  l(t)=\frac{R}{2} \sqrt{  \mu \int_0^t\frac{1}{\eta(u)} du } \quad ,
\end{equation}
leading to a total fluid volume of $V_p=\pi R^2 l(t)$ inside the pipe.
For reactive adhesives, whose hardening only depends on time, the
integral can be found by combining Eqs.~\ref{nuall} and
\ref{velocitypeneint} to
\begin{equation}\label{finallPUR}
  l(t)=\frac{R}{2} \sqrt{\mu} \left( \frac{\alpha t-\log(1+\gamma\exp^{\alpha t})}{\alpha \eta_g\exp^{\beta (1-C)}} \right)^{\frac{1}{2}} \quad .
\end{equation}
This way we obtain the time dependent penetration distance in a
straight single vessel, taking into account the applied pressure, the
capillarity effects, and the reactive hardening process. Note that for
adhesive types, whose viscosity changes when in contact with wood,
Eq.~\ref{velocitypeneint} can not be integrated so easily, since the
viscosity depends also on the concentration that changes with time.
Note that changes of the contact angle and furface tension of the
adhesives with solvent concentration are not considered in this work.

Since hard wood vessels are not straight, but weave tangentially
around rays, the penetration distance needs to be modified. Here we
simply describe vessels by the radius $R$, wavelength $\lambda$, and
amplitude $n$ (see Fig.~\ref{model}) of the oscillation in the
$z-y$-plane in the parametrized form as,
\begin{equation}\label{vesselpara}
  x^2+\frac{[y-n\cos(kz)]^2}{\sec^2[\arctan(n k\sin(kz))]} = R^2 \quad ,
\end{equation}
where $k=2\pi/\lambda$. By integrating the vessel length $l_v
$ along the $z$ direction, we obtain the volume 
\begin{equation}\label{vesselvolumen}
    V(l_v)=\pi R^2 l_v \left[1 + \frac{n^2 k^2}{4}\left( 1 + \frac{\sin(2kl_v)}{2kl_v}\right) \right] \quad.
\end{equation}
\subsection{Viscosity increase by diffusion}
Various adhesives contain solvents, whose concentration $C$ in the
mixture changes with time due to their diffusion into the cellular
structure through vessel walls. To take this effect into account, we
can write the solution of the diffusion equation in cylindrical
$z-r$-coordinates as
\begin{equation}\label{concentration}
  C(r,t)=\frac{C_0}{2Dt} \exp\left(-\frac{r^2}{4Dt }\right) \quad ,
\end{equation}
with the initial concentration of the solvent $C_0$ and its
diffusivity across the cell wall $D$. The average diffusivity of the
respective wood proved to be a good value. The mean value for the
solvent concentration inside the vessel follows as
\begin{equation}\label{concen}
  \overline{C} =  \int_0^R C(r,t)r dr = C_0 \left[1-\exp\left(-\frac{R^2}{4Dt} \right) \right] \quad .
\end{equation}
To obtain the complete equation for the evolution of the viscosity, we
go back to Eq.~\ref{nuall} and insert the concentration evolution
$\overline{C}$ into the respective concentration dependent parameters.
Note that we do not consider the diffusion of low molecular parts of
the adhesive. We also neglect the effect of swelling of the wood
skeleton due to moisture changes, since the size of vessels is rather
big compared to tracheids.
\subsection{Penetration into the network}
\begin{figure}
  \centering
  \includegraphics[scale=0.6]{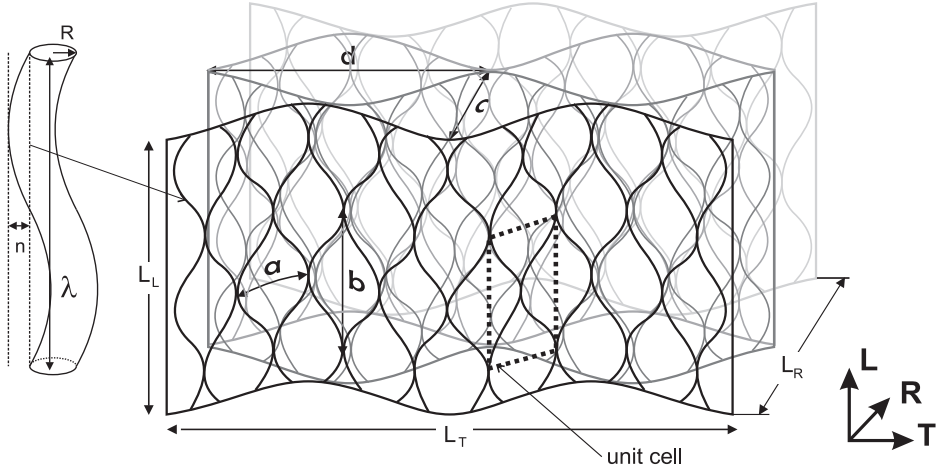}
  \caption{Vessel geometry and dimensions, vessel network, and unit
    cell. Shown are also the longitudinal ($L$), tangential ($T$), and
    radial ($R$) directions. \label{model}}
\end{figure}
The adhesive penetration is dominated by the flow inside the vessel
network, hence its topology determines the adhesive distribution. The
network is formed by bundles of vessels that divide and weave around
rays of various sizes. Inside the bundle, vessels interconnect by
contact zones when touching each other and can also interchange
positions \cite{philipp-09,bosshard-73}. Disorder in the network can
only be considered through a numerical approach. In order to be able
to derive an analytical model, we need to neglect disorder and use
average topological network parameters. We build up a regular network
using the average topological parameters $a$ and $b$ for connectivity
in tangential directions and $c$, $d$ for the connectivity in radial
direction. Fig.~\ref{model} shows the vessel network in three
dimensions with the geometrical parameters $a$, $b$, $c$ and $d$. Note
that $a$ and $b$ can be obtained from the size distribution of big and
middle sized rays that are mainly responsible for the splitting and
joining of the bundles of vessels. The parameters $c$ and $d$ however
are more difficult to obtain. Basically the probability for radial
network interconnections depends on the vessel density. We can
therefore find a relation between the vessel density and the parameter
$c$. $d$ however will remain a free parameter for transport in radial
direction. By separating the two geometric parameters $a$, $b$ and
$c$, $d$, we obtain anisotropic transport in the three principal
directions, longitudinal $L$, radial $R$, and tangential $T$ (see
Fig.~\ref{model}). Note that inclined samples with respect to the
principal axis can be considered after rotation. 

To describe the penetration process of adhesives into wood, we have to
define the bond line. The bondline is the whole region, where the
adhesive can be found. This includes the pure adhesive between the two
adherends and the area, where the adhesive has penetrated into the
wood structure. The adherends are two pieces of wood which have been
connected by the adhesive.  In our case, we will focus on the zone
where the adhesive layer and the adherent structure coexist.  The
procedure to obtain the maximum penetration depth is to calculate the
penetration separately in each principal direction (tangential $T$,
radial $R$, and longitudinal $L$, shown in Fig.~\ref{model}) and then
applying a rotation matrix to find the total penetration depth of the
adhesive when the growth ring angle and the angle between vertical and
longitudinal axis of the specimen are not zero.

\begin{figure}
  \centering
  \includegraphics[scale=0.7]{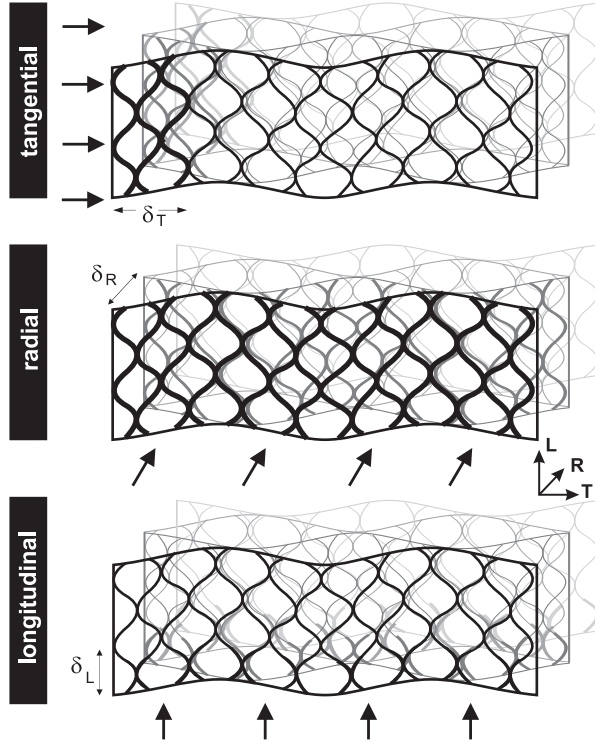}
  \caption{Adhesive penetration into the vessel network with respective depths in tangential ($\delta_T$), radial ($\delta_R$), and longitudinal ($\delta_L$) direction. Thick lines represent filled vessels.
    \label{modelpene}}
\end{figure}
Since we employ a regular network, a unit cell can be used that
consists of two single vessels with interconnections in the vertices
and the center of the cell (see Fig.~\ref{model}). Consequently we can
use Eq.~\ref{vesselvolumen} of the single vessel to obtain the volume
of the unit cell
\begin{equation}\label{cellvolumen}
  V_u=2A b g_T\quad \text{with} \quad g_T=1+\frac{a^2 \pi^2}{16 b^2}
\end{equation}
with the area $A=\pi R^2$. Because the unit cell can reproduce all the
network, we can use it to simplify the calculation of the penetration
depth on each direction and, using geometric properties, we can relate
the adhesive volume inside the network, the network parameters and the
penetration depth.
\subsubsection{Penetration in the tangential direction}
To consider the penetration in the \textit{tangential direction} (see
Fig.~\ref{modelpene}), we need to consider additionally to the tangential
waviness with wavelength $b$ and amplitude $a/4$ the radial waviness
with amplitude $c/4$ and wavelength $d$. Fig.~\ref{modelpene} shows how
the vessel network is filled by the adhesive. The path along one
radial wave as function of the tangential coordinate $x_T$ is given by
\begin{equation}\label{surroundlength}
  s_b = x_T \left[1+\frac{1}{2}\left(g_R-1 \right ) \left(1+\frac{d}{4\pi x_T}\sin \left(\frac{4\pi x_T}{d} \right) \right) \right]\quad \text{with} \quad g_R=1+\frac{c^2 \pi^2}{8 d^2}~~ .
\end{equation}
The number of layers accessible from the bond line is defined by
$2L_R/c$ with $L_R$ the sample width. Summing over all accessible
layers gives the total penetrated length as function of $x_T$
\begin{equation}\label{stotal}
  s_{total} = \frac{2 L_R}{c} s_b(x_T) \quad .
\end{equation}
To obtain the total penetrated volume we have to calculate the number
of unit cells $N_{uL}$ along $s_{total}$ and in longitudinal sample
direction $L_L$, this is $N_{uL}$=$\frac{s_{total}}{a}\frac{L_L}{b}$,
and then multiplied by the unit cell volume, Eq.~\ref{cellvolumen},
\begin{equation}\label{eq:vtot}
  V_T(x_T) = N_{uL} V_u \quad . 
\end{equation}
Assuming that the penetration of the adhesive is smaller than the
total wavelength $d$, $\frac{d\sin(4\pi x/d)}{4\pi x}$$\approx$$1$.
Therefore using Eqs.~\ref{cellvolumen} and \ref{stotal},
Eq.~\ref{eq:vtot} can be expressed as
\begin{equation}\label{totalvolume_T0}
  V_T(x_T) = x_T \frac{4A}{ac} L_R L_L g_T g_R \quad .
\end{equation}
When the adhesive stops to penetrate, this volume becomes the maximum
volumen $V$ inside the structure and the tangential coordinate $x_T$
transforms in the maximum penetration depth $\delta_T$,
\begin{equation}\label{depthT}
  \delta_T  = \frac{V a c}{4A L_R L_L g_T g_R} \quad . 
\end{equation}
\subsubsection{Penetration in the radial direction}
Following the idea of calculating the adhesive penetration in each
principal direction, the next step is to obtain the penetration depth
$\delta_R$ when the adhesive penetrates only in \textit{radial
  direction}.  Fig.~\ref{modelpene} illustrates the penetration of the
adhesive in order to relate the volume with the network parameter and
the penetration depth. We analogously count the volume occupied by
vessels as function of the radial coordinate $x_R$.  Again we
calculate the total length $s_{total}$ of the radial wave but now as
function of $x_R$ by $s_{total}$$=$$2 x_R/c s_b(L_T)$ and obtain the
number of unit cells
\begin{equation}\label{numbersection}
  N_{uR}(x_R) = \frac{2 x_R }{c}\frac{s_b(L_T)}{a}\frac{L_L}{b} \quad.
\end{equation}
The total volume occupied is given by multiplying the number of unit
cells $N_{uR}$ (Eq.~\ref{numbersection}) by the volume $V_u$ from
Eq.~\ref{cellvolumen}:
\begin{equation}\label{totalvolume_R0}
  V_T(x_R) = V_u N_{uR}=x_R \frac{4A}{ac} L_L s_b(L_T)g_T \quad .
\end{equation}
As before, we must now compare $V_T(x_R)$ with the volume occupied by
the adhesive $V$ with the maximum penetration depth in the radial
direction $x_R=\delta_R$ to obtain
\begin{equation}\label{totalvolume_R1}
  V = \delta_R \frac{4A}{ac} L_L s_b(L_T)g_T \quad ,  \quad\text{and} \quad \delta_R = \frac{V a c}{4A L_L s_b(L_T)g_T} \quad .
\end{equation}
Finally, we can insert the penetration path from
Eq.~\ref{surroundlength} and obtain
\begin{equation}\label{depthR}
  \delta_R = \frac{V a c}{4A L_L  L_T \left[1+\frac{1}{2}\left( g_R-1\right) \left(1+\frac{d\sin \left(\frac{4\pi L_T}{d} \right)}{4\pi L_T} \right) \right] g_T} \quad .
\end{equation}
\subsubsection{Penetration in the longitudinal direction}
We consider now the penetration only in \textit{longitudinal
  direction}. Fig.~\ref{modelpene} shows that the adhesive penetration
$\delta_L$ is basically along the vessels. This value is found again
by calculating the number of total unit cells, but now in the plane
$L_T L_R$, namely
\begin{equation}\label{numbercellunitsL}
  N_{u L} = \frac{2 L_R}{c} \frac{s_b(Lx)}{a} \quad .
\end{equation}
Multiplying $N_{u L}$ with the occupied volume $V(x_L)$ of the
adhesive for each vessel as function of $x_L$
(Eq.~\ref{vesselvolumen}), and taking into account that the
penetration $x_L$$<<$$b$, $\frac{b\sin(4\pi x_L/b)}{4\pi
  x_L}$$\approx$$1$, we obtain
\begin{equation}\label{numbervesselsL}
  V_T(x_L) = x_L \frac{4A}{ac} L_R s_b(L_T)\left( 2g_T-1\right) \quad .
\end{equation}
Again, comparing this volume with the adhesive volume, using
Eq.~\ref{surroundlength}, we can write the maximum penetration depth
as
\begin{equation}\label{depthL}
  \delta_L = \frac{V a c}{4A L_R  L_T \left[1+ \frac{1}{2}\left(g_R-1 \right)\left(1+\frac{d\sin \left(\frac{4\pi L_T}{d} \right)}{4\pi L_T} \right) \right] \left( 2g_T-1 \right)} \quad .
\end{equation}

Finally we obtained the maximum penetration depth in the three
principal directions (Eqs.~\ref{depthT},\ref{depthR},\ref{depthL}).
We can introduce the porosity $\epsilon$ of the wood which can be
extracted easily from experimental data \cite{philipp-09}. Expressing
the penetration depth in terms of porosity also simplifies the model
verification. The number of vessels $N_v$ in the plane $L_T L_R$
equals $N_v=2\cdot N_{uL}$. The porosity is therefore
\begin{equation}\label{porosity}
  \epsilon =\frac{N_vA}{L_R L_T}=\frac{4 A}{a c}\left[1 + \frac{1}{2}\left( g_R-1\right) \left(1+\frac{d\sin \left(\frac{4\pi L_T}{d} \right)}{4\pi L_T}\right)\right] \quad .
\end{equation}
Since porosity is a mean value, we can neglect the periodic part on
the right hand of the Eq.~\ref{porosity}. Inserting $\epsilon$ into
Eqs.~\ref{depthT}, \ref{depthR}, and \ref{depthL}, the maximum
penetration depths become
\begin{equation}\label{depthmaxp}
  \delta_R = \frac{V}{\epsilon L_L  L_T g_T} \quad , \delta_L = \frac{V}{\epsilon L_R  L_T \left( 2g_T-1 \right)} \quad ,  \delta_T  = \frac{V \left( g_R + 1\right) }{2\epsilon L_R L_Lg_T g_R} \quad . 
\end{equation}
\subsubsection{Limitation due to the total amount of applied adhesive}
Up to now, we calculated the penetration of an infinite amount of
non-hardening fluid. However the amount of applied adhesive and the
penetration time due to hardening are both limited. Therefore the
volume $V$ needs to be calculated considering these limitations. Both
limitations will lead to different penetrated volumes but only the
smaller one has a physical meaning. To calculate the volume with
penetration of hardening adhesives, we need to treat the
$L,R,T-$directions separately.

To consider \textit{tangential penetration} we employ
Eq.~\ref{depthmaxp} and apply adhesive only on the $RL$ plane. From
there, the adhesive can penetrate two channels with radius $R$ per
unit cell, and considering the number of unit cells on this face
$N_{uT}$, the volume penetrated after the hardening process, using
Eq.~\ref{velocitypeneint}, is
\begin{equation}\label{eq:vrl}
  V_{RL} = V_p \frac{L_L L_R}{c b} \quad .
\end{equation}
Inserting Eq.~\ref{eq:vrl} into Eq.~\ref{depthmaxp} and using
Eq.~\ref{porosity}, we obtain the penetration depth with hardening
$\delta_{T\rm h}$ as
\begin{equation}\label{depthTph}
  \delta_{T\rm h}  =  \frac{a}{b}\frac{l(t)}{g_T g_R} \quad. 
\end{equation}
For the \textit{radial penetration}, the penetrated volume is given
by,
\begin{equation}
  V_{LT} = 4V_p \frac{L_T L_L}{d a} \quad.
\end{equation}
By inserting $V_{LT}$ into Eq.~\ref{depthmaxp} and taking the mean
value of the periodic term, we obtain
\begin{equation}\label{depthRph}
  \delta_{R \rm h} = \frac{8}{\pi}\frac{a}{b}\frac{\sqrt{\frac{1}{2}\left( g_R-1\right)} l(t)}{\left( g_R + 1\right)g_T  } \quad .
\end{equation}
Finally, for the \textit{longitudinal penetration}, $\delta_{L \rm
  h}$, following a similar procedure the penetration depth becomes
\begin{equation}\label{depthmaxp2}
  \delta_{L \rm h} = \frac{l(t)}{\left(2g_T-1\right)} \quad .
\end{equation}

These values determine the maximum penetration depth that the adhesive
can reach until becoming solid. However it is possible, that not
enough adhesive is available, and penetration stops before. Using the
available adhesive volume $V$ in Eqs.~\ref{depthmaxp}, the penetration
depths $\delta_R$, $\delta_T$, and $\delta_L$ can be calculated and
compared to the hardening ones ($\delta_{R \rm h}$, $\delta_{T \rm
  h}$, $\delta_{L \rm h}$), e.g. if $\delta_R$$<$$\delta_{R \rm h}$,
to obtain the limiting case.
\subsection{Penetration depth for an arbitrary orientation}
In order to apply our model to real situations, we must have a way to
consider an orientation of the adhesive application surface that
deviates from the wood material system. Therefore we need to calculate
the global penetration depth $\delta_V$ and $\delta_{h}$ as function
of $\delta_R$, $\delta_L$, $\delta_{T}$, and $\delta_{R\rm h}$,
$\delta_{L\rm h}$, $\delta_{T\rm h}$, respectively. We can apply a
rotation matrix with the growth ring angle $\psi$ and the angle
$\theta$ between the vertical axis and the longitudinal axis of
specimen. Assuming that the adhesive is always applied on the $y-z$
plane, we apply two rotations in the principal coordinate system, one
in the radial direction and the other in the longitudinal direction
via the rotation matrix
\begin{equation}
  \underline{\underline{M}} = \left(
    \begin{array}{ccc}
      \sin(\psi) & -\cos(\theta) \cos(\psi) & -\cos(\psi) \sin(\theta) \\
      \cos(\psi) & \cos(\theta) \sin(\psi) & \sin(\theta) \sin(\psi) \\
      0 & -\sin(\theta) & \cos(\theta)
    \end{array}
  \right) \quad .
\end{equation}

Note that Eqs.~\ref{depthmaxp} give a dependence of the penetration
depths on the application areas $L_R\cdot L_T$, $L_L\cdot L_T$,
$L_L\cdot L_R$. We define a penetration vector $\underline{\delta} =
\delta_V \underline{S}$, where $\underline{S}$ is oriented normal to
the adhesive surface. $\underline{\delta}$ is given by
\begin{equation}
  \underline{\delta}=\frac{V}{\epsilon} \left(\frac{1}{g_T}, \frac{\frac{1}{2}\left( g_R+1 \right) }{g_T g_R}, \frac{1}{2g_T-1}  \right) ,
\end{equation}
in the principal coordinate system ($R-T-L$). Applying the rotation
matr ix $ \underline{\underline{M}}$ to the vector
$\underline{\delta}$, the $x$ component gives the maximum penetration
depth $\delta_V$
\begin{equation}\label{penetrationfinal}
  \delta_{V} = \frac{\Delta_T}{A_{ad}} \cos(\theta) \sin(\psi)+\frac{\Delta_L}{A_{ad}} \sin(\theta)\sin(\psi)+\frac{\Delta_R}{A_{ad}} \cos(\psi) ,
\end{equation}
with the area $A_{ad}$ of the surface where the adhesive is applied.
We can directly apply the rotation matrix to the vector
$\underline{\delta_{h}}$$=$$(\delta_{R\rm h}, \delta_{T \rm h},
\delta_{L \rm h})$ and find,
\begin{equation}\label{penetrationfinalh}
  \delta_h = \delta_{T\rm h} \cos(\theta) \sin(\psi)+\delta_{L \rm h}\sin(\theta)\sin(\psi)+\delta_{R \rm h} \cos(\psi) .
\end{equation}

With these derivations, we complete the geometric and dynamical
description of our model. The information about the dynamics of the
adhesive is included in the length $l(t)$ according to
Eq.~\ref{velocitypeneint}. Finally, our maximum penetration depth with
solvent diffusion can be calculated using Eq.~\ref{penetrationfinalh},
by replacing the concentration function in Eq.~\ref{nuall} for the
respective adhesives. In a next step we will apply the model to
experiments described in the first part of this paper \cite{part1}.
\section{Application of the Model}\label{applications}
Using synchrotron radiation X-ray tomographic microscopy (SRXTM) and
digital image analysis, we extracted bond lines from beech wood
samples, that were bonded with PUR, UF, and PVAC adhesives of
different viscosity under growth ring angles ranging from 0$^\circ$ to
90$^\circ$ in 15$^\circ$ steps \cite{part1}. Since our model is
periodic, we will calculate the maximum penetration depths for various
situations. The procedure is as follows: First we calculate the
penetration distance $l(t)$ of adhesive inside a single vessel. Note
that for PUR the calculation is without time dependence of the
concentration using Eq.~\ref{finallPUR}, while for PVAC and UF
adhesives with time dependence additionally Eq.~\ref{concen} is used.
The porosity $\epsilon$ and mean radius of the vessel $R$ are taken
from an earlier SRXTM study \cite{philipp-09} as $R$=28.03$\rm \mu$m
and porosity $\epsilon$=0.34. For all samples the mean applied
pressure was $P_A$$=$$0.7$MPa\cite{part1}. Literature values of the
surface tension $\sigma$
\cite{surfacetension1,surfacetension2,surfacetension3}, for the three
types of adhesives, are not large enough to compete with the applied
pressure term in Eqs.~\ref{velocitypene}, leading to negligible
capillarity effects, this means $\mu$$=$$P_A$. The parameters for the
viscosity $\eta_g$, $\alpha$, $\beta$ and $\gamma$ are taken from part
I \cite{part1}.
\begin{figure}[hbt] \centering
  \includegraphics[scale=0.65]{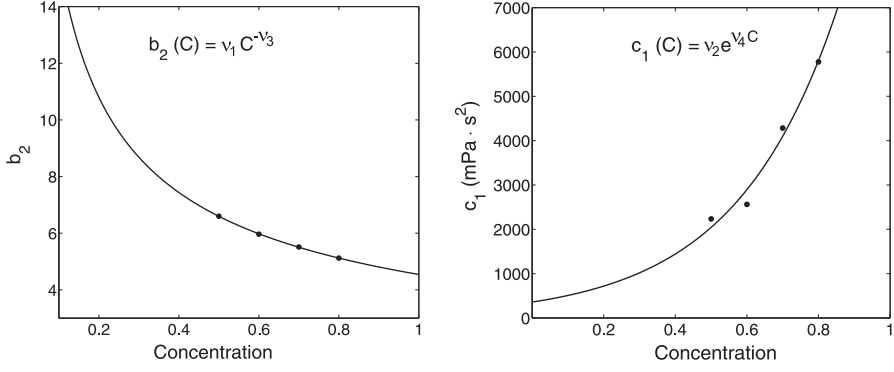}
  \caption{\label{viscosity} Dependence of the viscosity parameters
    $b_2$ and $c_1$ with the solvent concentration for UF adhesive.
    The dots denote the experimental data and the solid line the
    exponential fit. Experimental values for UF are
    $\nu_1=4.542$,$\nu_2=0.5379$,$\nu_3=358.3$,$\nu_4=3.482$.}
\end{figure}
\begin{itemize}
\item {For PUR adhesive, we choose the concentration $C=0.71$, and the
    parameters $\eta_g=4911\rm $mPa$\cdot$s, $\gamma$=9.74$\times$
    10$^{-5}$, $\alpha$=0.0028s$^{-1}$, $\beta$=0. The diffusion of
    solvent is not relevant. With these quantities the length for PUR
    adhesive $l(t)_{\rm PUR}$ is calculated using Eq.~\ref{finallPUR}
    to $l(t)_{\rm PUR}$=0.304m. This value seems huge at first sight,
    however it related to the path along the waving vessels that can
    easily reach lengths of 0.5m and above.}
\item {In the case of PVAC adhesive, the parameters are $C_0$=0.49,
    $\eta_g$=0.001859$\rm$ mPa$\cdot$s, $\gamma$=0, $\alpha$=0s$^{-1}$
    and $\beta$=29.64. The diffusivity of the solvent (water) for the
    samples is taken as $D$=3.0$\times$10$^{-9}\rm$ m$^2$ s$^{-1}$
    \cite{diffusivitydata1,diffusivitydata2}, and using
    Eq.~\ref{velocitypeneint}, we find a significantly lower value
    $l(t)_{\rm PVAC}$=0.7mm.}
\item {For UF we include the viscosity parameters $b_2$ and $c_1$ of
    Eq.~\ref{nuallUF} that change with the solvent concentration. We
    use the experimental viscosity data from Ref.\cite{part1} and fit
    it with analytical curves (see Fig.~\ref{viscosity}) to determine
    the concentration dependence of the viscosity parameters $b_2$ and
    $c_1$. After the identification of the right values for $b_2$ and
    $c_1$, we can integrate Eq.~\ref{velocitypeneint} and obtain a
    vessel penetration depth of $l(t)_{ \rm UF}$=1.1mm.}
\end{itemize}

The parameters $a$ and $b$ can be determined experimentally using
image processing (see Ref.\cite{part1}). In our case we measured the
area and the eccentricity of segmented rays and averaged over several
samples, and obtained values for $a$$=$$0.156$mm and $b$$=$$1.574$mm.
To eliminate variations due to the year ring structure, we used an
average porosity of $\epsilon$$=$$0.34$. The cylindric sample size had
$10$mm height and $3$mm diameter, leading to an adhesive area of
$A_{ad}$$=$$30 \rm mm^2$. As described in Ref.\cite{part1}, the
quantity of applied adhesive was around $200 \rm g/m^2$ for all
adhesives.

\begin{figure}[hbt]  \centering
  \includegraphics[scale=0.58]{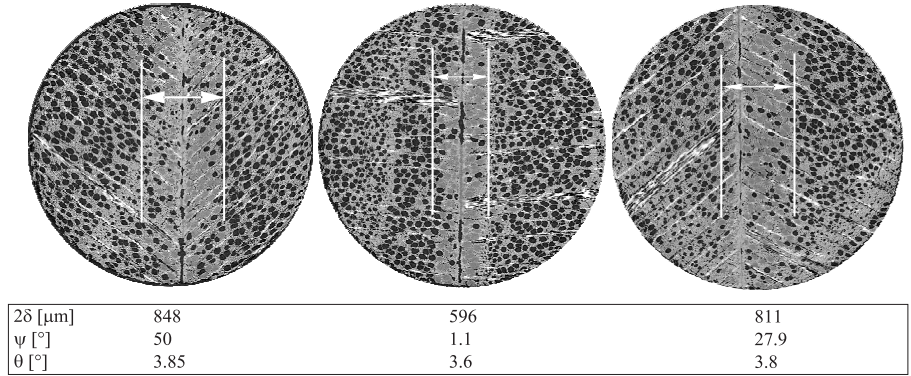}
  \caption{Bond lines with PUR adhesive in beech wood samples. The
    maximum penetration depth predicted by the model for all samples
    with growth ring angles $\psi$ and grain angle $\theta$ is shown
    by the white lines.\label{overview1}}
\end{figure}

\begin{figure}[hbt]  \centering
  \includegraphics[scale=0.75]{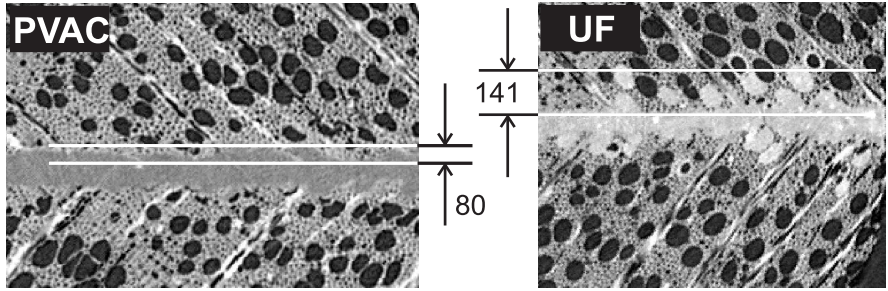}
  \caption{\label{overview2} Bond lines of PVAC and UF adhesive in
    beech wood with maximum predicted penetration depth for samples
    with the orientation angle $\psi$ and $\theta$. All dimensions are
    given in $\rm \mu$m.}
\end{figure}

\begin{figure}[hbt]  \centering
  \includegraphics[scale=0.75]{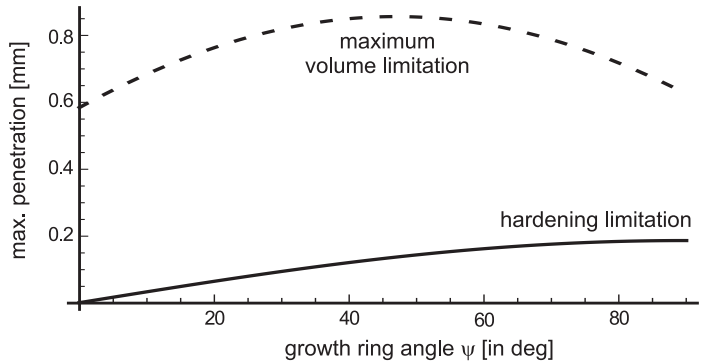}
  \caption{\label{figure7} Dependence of the penetration depth on the
    growth ring angle.}
\end{figure}

We compare the maximum penetration depth for samples with different
growth ring $\psi$ and grain angles $\theta$ (see
Figs.~\ref{overview1},\ref{overview2}). We fit the parameter $d$ to
obtain $g_R=1$, what can be interpreted as a lower probability of
interconnection in radial than in tangential direction.
\begin{itemize}
\item {For PUR adhesive, we choose a sample with angles,
    $\psi$$=$$50^\circ$ and $\theta$$=$$3.85^\circ$. If we calculate
    the maximum penetration depth using Eq.~\ref{penetrationfinal} we
    obtain a value of $\delta_h$$=$$38.5$mm with hardening as limiting
    factor, however using the volume limitation with
    Eq.~\ref{penetrationfinalh} we obtain $\delta_V$$=$$848\rm \mu m$.
    Therefore we can conclude that all the adhesive penetrated before
    hardening took place, leaving a starved bond line behind (see
    Fig.~\ref{overview1}). Note that the adhesive penetrates both wood
    pieces, but significantly deeper into on the application side
    (right side of samples in Fig.~\ref{overview1}). Therefore we have
    to take an average value of $\delta$$=$$424 \mu m$ showing good
    agreement with the experimental data. To test the model for other
    orientations $\psi,\theta$, we choose a sample with
    $\psi$$=$$1.1^\circ$ and $\theta$$=$$3.6^\circ$. This means we use
    the previous calculation but apply a new rotation matrix
    $\underline{\underline{M}}$. We obtain a penetration depth of
    $\delta_V$$=$$596 \rm \mu m$, $\delta$$=$$298 \rm \mu m$ and
    $\delta_h$$=$$1.32 \rm mm$. In Fig.~\ref{overview1} the quality of
    the analytical prediction is shown. We repeat this for angles
    $\psi$$=$$27.9^\circ$ and $\theta$$=$$3.8^\circ$ and obtain the
    penetration depths $\delta_V$$=$$811 \mu m$, $\delta$$=$$405.5 \rm
    \mu m$ and $\delta_h$$=$$23.6 \rm mm$ (compare
    Fig.~\ref{overview1}). These tests show that our model is a good
    approximation for the beech wood structure and therefore we fix
    the network parameters $a,b,c,d$ for further calculations.}
\item {We exemplify the penetration of PVAC using a sample oriented at
    angles $\psi$$=$$46.1^\circ$ and $\theta$$=$$3.4^\circ$ and
    calculate the maximum penetration depths from
    Eqs.~\ref{penetrationfinal} and \ref{penetrationfinalh}. We find
    $\delta_V$$=$$851 \rm \mu m$, leading to $\delta$$=$$425.5 \rm \mu
    m$ and $\delta_h$$=$$80 \rm \mu m$. Therefore the maximum
    penetration depth is limited by the hardening process. In
    Fig.~\ref{overview2}, we show that almost all the adhesive remains
    in the bond line with only a small quantity of adhesive inside the
    vessel network.}
\item {For UF we repeat the same procedure as before on a sample with
    orientation angles $\psi$$=$$37.2^\circ$ and
    $\theta$$=$$4.3^\circ$. We find that the penetration depths are
    $\delta_V$$=$$856 \rm \mu m$, $\delta$$=$$428 \rm \mu m$, and
    $\delta_h$$=$$141 \rm \mu m$ and again the penetration of the
    adhesive is limited by adhesive hardening. Fig.~\ref{overview2}
    shows the sample with the predicted penetration depth, exhibiting
    excellent agreements between the analytical prediction and the
    experiments.}
\end{itemize}
To study the dependence of maximum penetration depth on the growth
ring angle, we take the values for UF and keep all parameters fixed,
except the growth ring angle. In Fig.~\ref{figure7} we see the two
limiting conditions for the penetration depth. The penetration depth
is an increasing function of the growth ring angle for the hardening
limitation case, and we observe a distinct maximum at approximately
$48$$^\circ$ in the case when the maximum available volume is the
limitation. This observation is in agreement with PUR adhesive that
fulfills the volume limiting condition, as demonstrated in part I of
this work \cite{part1}. This result shows that even though we reduce
the wood anatomy to a homogeneous, regular network, adhesive
transport, the beech wood seems well described by the model and for a
desired penetration depth, the model can predict the optimal growth
ring angle of the samples.

\begin{figure}[hbt]  \centering
  \includegraphics[scale=0.145]{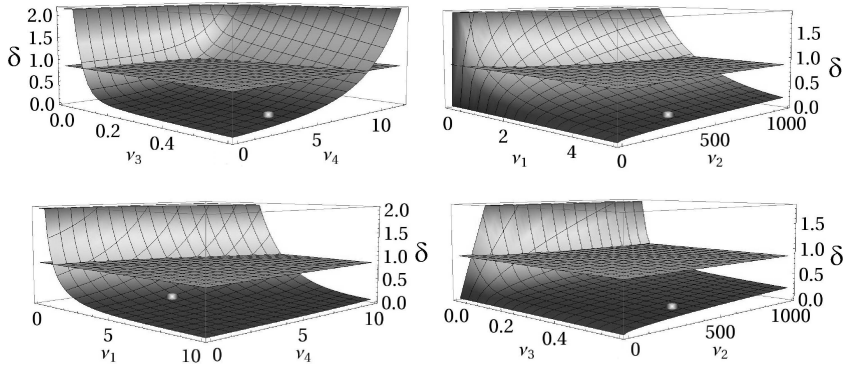}
  \caption{\label{optim} Penetration depth $\delta$ in mm as function
    of adhesive parameters from Fig.~\ref{viscosity} for UF and growth
    ring angle of 45$^\circ$. The experimentally obtained value for UF
    is marked with the white dot. (Color version online)}
\end{figure}

\begin{figure}[hbt]  \centering
  \includegraphics[scale=0.5]{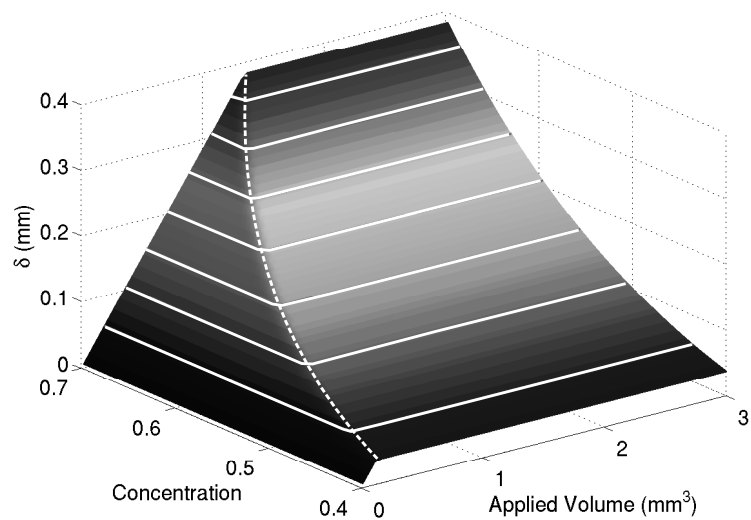}
  \caption{\label{peneoptim} Penetration depth $\delta$ in mm as
    function of solvent concentration and total amount of applied
    adhesive for UF and growth ring angle of 45$^\circ$. The solid
    lines represent the fixed $\delta$ and the dashed line divides the
    surface into two regions, the right one when the penetration is
    limited by the hardening process and the left one, when the total
    volume of applied adhesive is the limiting factor. (Color version
    online)}
\end{figure}

Our model can also be used to design new adhesives with optimized
properties, like reactivity, if an ideal penetration depth is to be
reached. Fig.~\ref{optim} shows the maximum penetration depth for a
wide range of adhesive parameters $\nu_{1}, ..., \nu_{4}$ from
Fig.~\ref{viscosity}.  We show this in four plots combining two
parameters. Horizontal planes represent the case where all available
adhesive is inside the vessel structure, while the curved surfaces
show the penetration limit due to adhesive hardening. The intersection
line (see Fig.~\ref{optim}) separates regions with complete
penetration from those, where penetration is limited by adhesive
hardening.  Therefore, Fig.~\ref{optim} allows to choose a pair of
reactivity parameters in order to obtain a desired penetration depth.
The model can also be used to minimize solvent concentration and
amount of applied adhesive for a required penetration depth.
Fig.~\ref{peneoptim} illustrates the maximum penetration depth as
function of the solvent concentration and the total amount of applied
adhesive. The solid lines represent the proportions between solvent
concentration and the total applied volume of adhesive which give the
same penetration depth.

\section{Discussions and Conclusions}\label{conclusions}
We presented an analytical model for the prediction of the penetration
depth of adhesives, paint or hardening fluids in general, into the
beech wood structure. Since we focused on hard wood, the pore space of
wood is formed by a network of interconnected vessels. The network is
characterized by parameters that are related to amplitude and
wavelength of the oscillating vessels in tangential and radial
direction and the porosity of wood originating from the vessel
network. We compared the model to experiments and found good agreement
for various adhesives. Therefore, even though we reduce the wood
anatomy to a homogeneous, regular network, adhesive transport for the
much more disordered, complex pore space of real beech wood seems well
described.

The analytical model considers generic types of adhesive hardening.
However if other special fluids are to be considered, only the
viscosity dependence on concentration and time need to be known.  We
applied the model to three major types of adhesive which are PUR, UF,
and PVAC and compared the respective penetration depths. For adhesives
whose hardening process depends on the change of concentration, we
include a description for solvent concentration diffusion inside the
beech wood. This approach makes the model applicable for adhesives
like UF and PVAC. Penetration is limited by two things: the
penetration due to the applied pressure that is arrested by hardening
processes, and the total amount of applied adhesive that is available
to penetrate into the vessel network. The smaller penetration depth is
the limiting one. By comparing the model with the experimental data we
showed that it is possible to model the maximum penetration depth for
three different used adhesives, namely PUR, PVAC, and UF.

Our model is sufficiently simple to allow for a broad applicability.
By determining the morphological and rheological parameters, it can be
applied to a wide range of wood species and to fluids with various
hardening kinematics to predict the penetration depth of these fluids
into porous structures, when transport is dominated by capillarity.
\section*{Acknowledgement}
The authors are grateful for the financial support of the Swiss
National Science Foundation (SNF) under Grant No. 116052.


\begin{thebibliography}{}
\bibitem{marra-92} A.A. Marra, Technology of wood bonding, Van Nostrand Reinhold, New York, NY (1992).
\bibitem{Custodio-etal-2009} J. Custodio, J. Broughton, H. Cruz, A review of factors influencing the durability of structural bonded timber joints, International Journal of Adhesion and Adhesives, 29, 173-185 (2009).
\bibitem{wang-yan-2005} W.Q. Wang, N. Yan, Characterizing liquid resin penetration in wood using a mercury intrusion porosimeter, Wood and Fiber Science, 37, 505-514 (2005).
\bibitem{kamke-lee-2007} F.A. Kamke, J.N. Lee, Adhesive Penetration of Wood - A Review, Wood and Fiber Science, 39(2), 205-220 (2007).
\bibitem{siau-84} J.F. Siau, Transport processes in wood. Springer, New York, NY (1984).
\bibitem{philipp-09} P. Hass, F.K. Wittel, S.A. McDonald, F. Marone, M. Stampanoni, H.J. Herrmann, P. Niemz,  Pore space analysis of beech wood - the vessel network. Submitted to Holzforschung. Preprint visible in electronic form.
\bibitem{sernek-resnik-kamke-99} M. Sernek, J. Resnik, and F.A. Kamke, Penetration of liquid Urea-Formaldehyde Adhesive into Beech Wood, Wood and Fiber Science, 31(1), 41-48 (1999).
\bibitem{niemz-etal-2004} P. Niemz, D. Mannes, E. Lehmann, P. Vontobel, and S. Haase, Untersuchungen zur Verteilung des Klebstoffes im Bereich der Leimfugen mittels Neutronenradiographie und Mikroskopie, European Journal of Wood and Wood Products 62, 424-432 (2004).
\bibitem{collett-72} B.M. Collett, A Review of Surface and Interfacial Adhesion in Wood Science and Related Fields, Wood Science and Technology, 6,1-42 (1972).
\bibitem{suchsland-58} O. Suchsland, \"Uber das Eindringen des Leims bei der Holzverleimung und die Bedeutung der Eindringtiefe für die Fugenfestigkeit, European Journal of Wood and Wood Products 16(39),101-108 (1958).
\bibitem{part1} P. Hass, M. Mendoza, F.K. Wittel, P. Niemz, H.J. Herrmann, Adhesive Penetration of Hard Wood: Part I: Experiments on Beech, Submitted to Wood Science and Technology. Preprint visible in electronic form.
\bibitem{washburn-21} E.W. Washburn, The Dynamics of Capillarity Flow, Phys. Rev. 17, 273-283 (1921).
\bibitem{bosshard-73} H.H. Bosshard, L. Kucera, The Network of Vessel System in Fagus sylvatica L., European Journal of Wood and Wood Products 31,437-445 (1973).
\bibitem{surfacetension1} A. Bhattacharya and P. Ray, Studies on Surface Tension of Poly(Vinyl Alcohol): Effect of Concentration, Temperature, and Addition of Chaotropic Agents, J. of Appl. Poly. Science, 93, 122-130 (2004).
\bibitem{surfacetension2} C.-Y. Hse, Surface tension of phenol-formaldehyde wood adhesives, Holzforschung 26, 82-85 (1972).
\bibitem{surfacetension3} S. Lee, T.F. Shupe, L.H. Groom and C.Y. Hse, Wetting Behaviors of Phenol- and Urea-Formaldehyde Resins as Compatibilizers, Wood and Fiber Science 39, 482-492 (2007).
\bibitem{diffusivitydata1} S. Kurjatko and J. Kudela, Wood Structure and Properties, Arbora Publisher (1998).
\bibitem{diffusivitydata2} W. Olek, P. Perr{\'e} and J. Weres, Inverse analysis of the transient bound water diffusion in wood, Holzforschung 59, 38-45 (2005).   
\end{thebibliography}
\end{document}